\documentclass[prd,aps,nofootinbib,onecolumn,showpacs,10pt]{revtex4-1}
\usepackage{graphicx,epsfig}

\usepackage{epsf}
\usepackage{graphicx}
\def\ds{\displaystyle}

\begin{document}
\def\ds{\displaystyle}
\def\beq{\begin{equation}}
\def\eeq{\end{equation}}
\def\bea{\begin{eqnarray}}
\def\eea{\end{eqnarray}}
\def\beeq{\begin{eqnarray}}
\def\eeeq{\end{eqnarray}}
\def\ve{\vert}
\def\vel{\left|}
\def\ver{\right|}
\def\nnb{\nonumber}
\def\ga{\left(}
\def\dr{\right)}
\def\aga{\left\{}
\def\adr{\right\}}
\def\lla{\left<}
\def\rra{\right>}
\def\rar{\rightarrow}
\def\nnb{\nonumber}
\def\la{\langle}
\def\ra{\rangle}
\def\bdll{$B_c \rightarrow D_{q'}^*l^+ l^-$}
\def \xjga{$\chi_{c0} \rar J/\psi \gamma $~}
\def \xUga{$\chi_{b0} \rar \Upsilon \gamma $~}
\def\ba{\begin{array}}
\def\ea{\end{array}}
\title{  \bf   Strong coupling constant of h$_{b}$ vector to the pseudoscalar and vector
$B_{c}$\\mesons in QCD sum rules}
\author{V. Bashiry$^{\dag1}$, A. Abbasi$^{\ast2}$}
\affiliation{$^{\dag}$Cyprus International University, Faculty of Engineering, Nicosia,
Northern Cyprus, Mersin 10, Turkey}
\affiliation{$^{\ast}$Eastern Mediterranean University,Department of Physics, G. Magusa,
North Cyprus, Mersin-10, Turkey}
\affiliation{$^{1}$E-mail:bashiry@ciu.edu.tr, $^{2}$E-mail: akbar.abbasi@emu.edu.tr}

\begin{abstract}
The strong coupling constant g$_{h_{b}B_{c}^{PS}B_{c}^{V}}$ is calculated 
using the three-point QCD sum rules method. We use correlation functions to
obtain these strong coupling constants with contributions of both B$_{c}^{PS} $ and B$_{c}^{V}$mesons as off-shell states. The contributions of two gluon condensates as a radiative correction are considered. The results show that g$_{h_{b}B_{c}^{PS}B_{c}^{V}}=8.80\pm 2.84 GeV^{-1}$and g$_{h_{b}B_{c}^{V} B_{c}^{PS}}=9.34\pm 3.12 GeV^{-1}$ in the  $B_{c}^{PS}$ and $B_{c}^{V}$ off-shell state, respectively.

\end{abstract}
\pacs{ 11.55.Hx,  13.75.Lb, 13.25.Ft,  13.25.Hw}

\maketitle

\section{Introduction}
 Measurements of  masses, total widths and transition
rates of heavy quark bound states  serve as important benchmarks for the predictions of QCD-inspired
potential models, non-relativistic QCD, lattice QCD and QCD sum
rules\cite{Aubert}.
The $h_b$ mesons are bound states of $\overline{b}b$ quarks. The system is approximately non-relativistic due to the large $b$ quark mass,
and therefore the quark-antiquark QCD potential can be investigated via
$\bar{b}b$ spectroscopy. These mesons are intermediate states between $Y(3S)$ to $\eta_{b}(1S)$ with the processes $Y(3S)\rightarrow\pi^{+}\pi^{-}\pi^{0}h_{b}$ and decay to ground state $\gamma\eta_{b}$. The  $h_{b}(1P)$ state is spin-singlet P-wave bound state of $b\overline{b}$ quarks which was observed for the first time by Belle collaboration with significance of $5.5\sigma$\cite{Belle}. It has been conjectured that this meson often decays into an intermediate two-body states of $B$ mesons, then undergoes final state interactions.  This meson ($h_b(1P)$) is used to study of the P-wave spin-spin (or hyperfine) interaction. Therefore, theoretical calculations on the physical parameters of this meson and their comparison with experimental data should give valuable  information as regards  the nature of hyperfine interaction. However, most of the theoretical studies deal with the non-perturbative QCD calculations.
The mass and leptonic decay constant of $h_b(1P)$ mesons have been calculated \cite{Bashiry}. These physical parameters help us to calculate the other physical parameters, i.e., the rates  of various decay modes and coupling constants.

In this work, we evaluate the strong coupling constant, $g_{h_{b}B_{c}%
^{PS}B_{c}^{V}}$ within the framework of three-point QCD sum rules. We consider
contributions of both $B_{c}^{V}$ and $B_{c}^{PS}$ mesons as off-shell states. The contributions of the bare loop diagram and
the two-gluon condensate diagrams as radiative corrections are evaluated. We assume that $h_b$ is on-shell, that may decay to the intermediate $B_{c}^{V}$ and $B_{c}^{PS}$ mesons. In this regard, the coupling
constants help us to describe the intermediate state of two-body decay of
meson into B$_{c}^{V}$ and B$_{c}^{PS}$ mesons when one of these mesons is off-shell. The intermediate states decay
into the final states with the exchange of virtual mesons. Indeed, without
understanding the mechanism of intermediate states, we are not able to analyze
the results of ongoing experiments properly.

Here, we use the same technique
for the study of the couplings such as  $D^{\ast}D_{s}K$, $D_{s}^{\ast}DK$ \cite{Bracco1,Wang1}, $D_{0}D_{s}K$, $D_{s_{0}}DK$ \cite{Rodrigues},
$D^{\ast}D\pi $ \cite{Navarra} , $D_{s}D^{\ast}K,D_{s}^{\ast}DK$
\cite{Bracco2}, $B_{s_{0}}BK$, $B_{s_{1}}B^{\ast}K$ \cite{Wang2},
$D_{s}^{\star}DK^{\ast}$\cite{Azizi} and $\eta_{b}BB^{\ast}$vertex from QCD
sum rule\cite{Cui}, $B_s^{\ast}B K$\cite{ref1}, $B_{1s} B^{\ast}K$\cite{ref2}, $B^{\ast}B^{\ast} \rho$\cite{ref2} .

The present work is organized as follows:  In
section II, we introduce the QCD sum rules technique
where analytical expressions of the $g_{h_bB_{c}^{V}B_{c}^{PS}}$ strong coupling constant are obtained.
Section III is devoted to the numerical analysis and discussion.

\section{QCD sum rules for the form factors}
In this section, we present QCD sum rules calculation for the form factor of
the $h_{b}B_{c}^{V}B_{c}^{PS}$ vertex. The three-point correlation function
associated with the $h_b B_{c}^{V}B_{c}^{PS}$ vertex is given by%

\begin{equation}\label{CorrelationFuncPhys1}
\Pi_{\mu\nu}^{B_{c}^{PS}}(p^{\prime},q)=i^{2}\int d^{4}xd^{4}ye^{ip^{\prime
}.x}e^{iq.y}\langle0|T(j_{\nu}^{B_{c}^{V}}(x)j^{B_{c}^{PS}}(y)j_{\mu}^{h_{b}%
}(0))\left\vert 0\right\rangle ,\label{1n}
\end{equation}
where, the $B_{c}^{PS}$is off-shell state, and:
\begin{equation}
\Pi_{\mu\nu}^{B_{c}^{V}}(p^{\prime},q)=i^{2}\int d^{4}xd^{4}ye^{ip^{\prime}%
.x}e^{iq.y}\langle0|T(j^{B_{c}^{PS}}(x)j_{\nu}^{B_{c}^{V}}(y)j_{\mu}^{h_{b}%
}(0))\left\vert 0\right\rangle ,\label{2n}%
\end{equation}
where the $B_{c}^{V}$ is off-shell state, $q$ is transferred
momentum, and $T$ is the time ordering operator.

We describe each meson field in terms of the quark field operators as follows:
\begin{eqnarray}
j_{\nu}^{B_{c}^{V}}(x)  & =&\overline{c}(x)\gamma_{\nu}b(x)\label{3n}\\
j^{B_{c}^{PS}}(y)  & =&\overline{c}(y)\gamma_{5}b(y)\nonumber\\
j_{\mu}^{h_{b}}(0)  & =&\overline{b}(0)\gamma_{\mu}\gamma_{5}b(0)\nonumber
\end{eqnarray}
 The above correlation functions need to be calculated in two
different ways: on the theoretical side, they are evaluated with the help of the
operator-product expansion (OPE), where the short and large-distance effects
are separated;  on the phenomenological side, they are calculated in terms of
hadronic parameters such as masses, leptonic decay constants, and form
factors. Finally, we aim to equate structures of the two representations.

Performing the integration over $x$ and $y$ of Eq. (\ref{CorrelationFuncPhys1}) we get:

\begin{equation}\label{corr}
\Pi_{\mu\nu}^{B_{c}^{PS}}(p^{\prime},q)=\frac{\left\langle 0\right\vert
j_{\nu}^{B_{c}^{V}}\left\vert B_{c}^{V}(p^{\prime},\epsilon^{\prime
})\right\rangle \left\langle 0\right\vert j^{B_{c}^{PS}}\left\vert B_{c}
^{PS}(q)\right\rangle \left\langle B_{c}^{V}(p^{\prime},\epsilon^{\prime
})\right\vert B_{c}^{PS}(q)\left\vert h_{b}(p,\epsilon)\right\rangle
\left\langle h_{b}(p,\epsilon)\right\vert j_{\mu}^{h_{b}}\left\vert
0\right\rangle }{(q^{2}-m_{B_{c}^{PS}}^{2})(p^{2}-m_{h_{b}}^{2})(p^{\prime
^{2}}-m_{B_{c}^{V}}^{2})}+...
\end{equation}
In order to finalize  the calculation of the phenomenological side, it is necessary to
know the effective Lagrangian  for the interaction of the  the vertex $h_b B_c^V B_c^{PS}$, which is given as follows:

\begin{equation}\label{Lag}
  {\cal{L}}=g_{h_{b}B^VB^{PS}}B^{PS}\{ (\partial_{\alpha}h^{\sigma})(\partial^{\alpha}B^V_{\sigma})-(\partial^{\beta}h_{\alpha})(\partial^{\alpha}B^V_{\beta}) \}
\end{equation}
where $h$ is axial-vector meson field($h_b(1P)$ field), $B^V$ is the vector meson field and $B^{PS}$ is the pseudoscalar meson field.

The matrix elements of the Eq. (\ref{corr}) can be related to the hardronic
parameters as follows:%

\begin{eqnarray}\label{def}
\left\langle 0\right\vert j_{\nu}^{B_{c}^{V}}\left\vert B_{c}^{V}(p^{\prime
},\epsilon^{\prime})\right\rangle  & =&m_{B_{c}^{V}}f_{B_{c}^{V}}\epsilon_{\nu
}^{\prime}\nonumber \\
\left\langle 0\right\vert j^{B_{c}^{PS}}\left\vert B_{c}^{PS}(q)\right\rangle
& =&i\frac{m_{B_{c}^{PS}}^{2}}{m_{b}+m_{c}}f_{B_{c}^{PS}}\\
\left\langle B_{c}^{V}(p^{\prime},\epsilon^{\prime})\right\vert B_{c}
^{PS}(q)\left\vert h_{b}(p,\epsilon)\right\rangle  & =&ig^{B_c^{PS}}_{h_{b}BB}
[(p.p^{\prime})(\epsilon.\epsilon^{\ast\prime})-(p.\epsilon^{\ast\prime
})(p^{\prime}.\epsilon)]\nonumber\\
\left\langle h_{b}(p,\epsilon)\right\vert j_{\mu}^{h_{b}}\left\vert
0\right\rangle  &=&m_{h_{b}}f_{_{h_{b}}}\epsilon_{\mu}^{\ast}\nonumber
\end{eqnarray}
where; $g_{h_{b}BB}$ is strong coupling constant when $B_{c}^{PS}$ is
off-shell and $\epsilon$ and $\epsilon^{\prime}$ are the polarization vectors
associated with the $h_{b}$ and $B_{c}^{V}$ respectively. Substituting Eq. (\ref{def}) in
Eq. (\ref{corr}) and using the summation over polarization vectors via,

\begin{equation}\label{}
\epsilon_{\nu}\epsilon_{\theta}^{\ast}=-g_{\nu\theta}+\frac{q_{\nu}q_{\theta}
}{m_{B_{c}^{PS}}^{2}},
\end{equation}
\begin{equation}
\bigskip\epsilon_{j}^{^{\prime}}\epsilon_{\mu}^{^{\prime}\ast}=-g_{j\mu}%
+\frac{p_{j}p_{\mu}}{m_{B_{c}^{V}}^{2}},
\end{equation}
the phenomenological or physical side for $B_{c}^{V}%
$off-shell result is found to be:
%
\begin{equation}
\Pi_{\mu\nu}^{B_{c}^{V}}(p^{\prime},q)=-g^{B_{c}^{V}}_{h_{b}BB}(q^{2})\frac{m_{B_{c}^{V}}f_{B_{c}^{V}}\frac{m_{B_{c}^{PS}}^{2}}{m_{b}+m_{c}}f_{B_{c}^{PS}}m_{h_{b}%
}f_{_{h_{b}}}}{(q^{2}-m_{B_{c}^{V}}^{2})(p^{2}-m_{h_{b}}%
^{2})(p^{\prime^{2}}-m_{B_{c}^{PS}}^{2})}(p.p^{\prime})g_{\mu\nu}+...
\end{equation}

and "..." represents the
contribution of the higher states and continuum.

We  compare the coefficient of the $(p.p^{\prime})g_{\mu\nu}$ structure for further calculation from different approaches of the correlation functions.

Also, a similar expression of the physical side of the correlation function
for $B_{c}^{PS}$off-shell meson is the following:

\begin{equation}
\Pi_{\mu\nu}^{B_{c}^{PS}}(p^{\prime},q)=-g^{B_{c}^{PS}}_{h_{b}BB}(q^{2})\frac{m_{B_{c}^{V}
}f_{B_{c}^{V}}\frac{m_{B_{c}^{PS}}^{2}}{m_{b}+m_{c}}f_{B_{c}^{PS}}m_{h_{b}
}f_{_{h_{b}}}}{(q^{2}-m_{B_{c}^{PS}}^{2})(p^{2}-m_{h_{b}}
^{2})(p^{\prime^{2}}-m_{B_{c}^{V}}^{2})}(p.p^{\prime})g_{\mu\nu}+...
\end{equation}

 In the following, we calculate the correlation functions on the QCD
side using the deep Euclidean space ($p^{2}\rightarrow-\infty$ and
$p^{\prime2}\rightarrow-\infty$). Each invariant amplitude $\Pi_{\mu\nu}%
^{_{i}}(p^{\prime},q)$ where $i$ stands for $B_{c}^{PS}$or $B_{c}^{V}$
consists of perturbative (bare loop, see Fig. 1), and non-perturbative parts
(the contributions of two-gluon condensate diagrams, see Fig. (2) )as:%

\begin{equation}
\Pi_{\mu\nu}^{_{i}}(p^{\prime},q)=(\Pi_{per}+\Pi_{nonper})(p.p^{\prime})g_{\mu\nu}%
\end{equation}

The perturbative contribution and gluon condensate contribution can be defined
in terms of double dispersion integral as

\begin{equation}
\Pi_{per}=-\frac{1}{4\pi^{2}}\int ds^{\prime}\int ds\frac{\rho(s,s^{\prime
},q^{2})}{(s-p^{2})(s^{\prime}-p^{\prime2})}+\text{subtractionterms},
\end{equation}
where, $\rho(s,s^{\prime},q^{2})$ is the spectral density. It is aimed to
evaluate the spectral density by considering the bare loop diagrams (a)
and (b) in Fig. 1 for $B_{c}^{V}$ and $B_{c}^{PS}$off-shell, respectively. We
use the Cutkosky method to calculate these bare loop diagrams and replace the
quark propagators of Feynman integrals with the Dirac Delta Function:

\begin{equation}
\frac{1}{q^{2}-m^{2}}\rightarrow(-2\pi i)\delta(q^{2}-m^{2}).
\end{equation}
Results of spectral density are found to be:
\begin{eqnarray}
\rho^{B_{c}^{PS}}(s,s^{\prime},q^{2})  &=&\frac{N_{c}}{\lambda^{3/2}
(s,s^{\prime},q^{2})}\{2 m_b^2 m_c s - m_c s (2 mc^2 + q^2 + s -s^\prime) \nonumber\\&+& 2 m_b^3 (q^2 + s^\prime) -
 m_b (2 m_c^2 + q^2 + s - s^\prime) (q^2 + s^\prime))\},
\end{eqnarray}
and:
\begin{eqnarray}
\rho^{B_{c}^{V}}(s,s^{\prime},q^{2})  &=&\frac{N_{c}}{\lambda^{3/2}%
(s,s^{\prime},q^{2})}\{2 m_b^2 m_c s - m_c s (2 mc^2 + q^2 + s -s^\prime) \nonumber\\&+& 2 m_b^3 (q^2 + s^\prime) -
 m_b (2 m_c^2 + q^2 + s - s^\prime) (q^2 + s^\prime))\},
\end{eqnarray}
\begin{figure}[h!]
\begin{center}
\includegraphics[width=12cm]{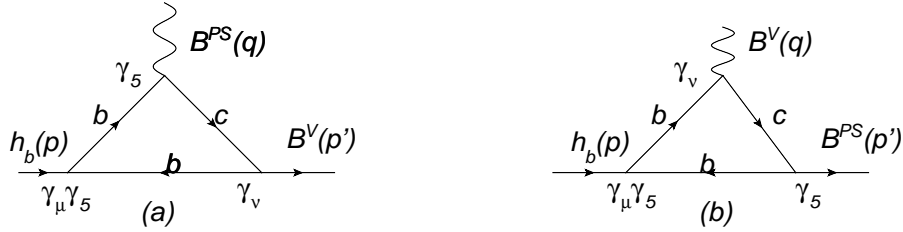}
\end{center}
\caption{(a) and (b): Bare loop diagram for the $B_c^{PS}$  and
$B_c^{V}$ off-shell, respectively; } \label{bare}
\end{figure}
where $\lambda(a,b,c)=a^{2}+b^{2}+c^{2}-2ac-2bc-2ab$ and the color number $N_{c}=3$.
 The physical region in $s$ and $s^{\prime}$ plane is described
by the following inequality:%

\begin{equation}
-1\leq f^{i}(s,s^{\prime})=\frac{s(-2m_{b}^{2}+2m_{c}^{2}+q^{2}+s-s^{\prime}%
)}{\lambda^{1/2}(m_{b}^{2},m_{b}^{2},s)\lambda^{1/2}(s,s,q^{2})}\leq1.
\end{equation}
where $i$ indicates two states of $B_{c}^{PS}$and $B_{c}^{V}$ off-shell
meson.

The diagrams for the contribution of the gluon condensate in the case
$B_{c}^{PS}$ off-shell are depicted in (a), (b), (c), (d), (e) and (f) in Fig. (2).
\begin{figure}[h!]
\begin{center}
\includegraphics[width=12cm]{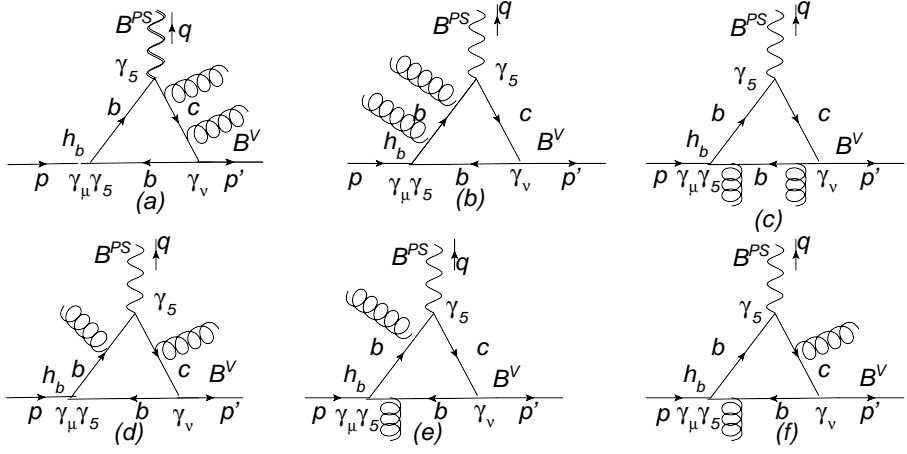}
\end{center}
\caption{Two-gluon condensate diagram as a radiative corrections  for the $B_c^{PS}$
off-shell; } \label{gluon}
\end{figure}
All diagrams are calculated in the Fock-Schwinger fixed-point
gauge\cite{Fock,Schwinger,Smilga} where we assume $x^{\mu}A_{\mu}^{a}=0$ for the gluon field $A_{\mu}^{a}$. Then, the vacuum gluon field is %

\begin{equation}
A_{\mu}^{a}(k^{\prime})=-\frac{i}{2}(2\pi)^{4}G_{\rho\mu}^{a}(0)\frac
{\partial}{\partial k_{\rho}^{\prime}}\delta^{(4)}(k^{\prime}),
\end{equation}
where $k^{\prime}$is the gluon momentum.

In this calculation, we need to solve
the following two types of integrals:
\bea \label{e7323} I_0[a,b,c] = \int
\frac{d^4k}{(2 \pi)^4} \frac{1}{\left[ k^2-m_b^2 \right]^a \left[
(p+k)^2-m_b^2 \right]^b \left[ (p^\prime+k)^2-m_c^2\right]^c}~,
\nnb \\ \nnb \\
I_\mu[a,b,c] = \int \frac{d^4k}{(2 \pi)^4} \frac{k_\mu}{\left[
k^2-m_b^2 \right]^a \left[ (p+k)^2-m_b^2 \right]^b \left[
(p^\prime+k)^2-m_c^2\right]^c}~,
 \eea
where $k$ is the
momentum of the spectator quark $b$. These integrals can be
calculated by flipping  to Euclidean space--time and using
Schwinger representation for the Euclidean propagator

 \bea
\label{e7324} \frac{1}{(k^2+m^2)^n} = \frac{1}{\Gamma(n)}
\int_0^\infty d\alpha \, \alpha^{n-1} e^{-\alpha(k^2+m^2)}~,
 \eea

the Borel transformation is as follows:

\begin{equation}
B_{\widehat{p}}(M^{2})e^{-\alpha p^{2}}=\delta(\frac{1}{M^{2}}-\alpha).
\end{equation}
where $M$ is Borel parameter.

 We integrate over loop momentum
and  two parameters that we have used in the exponential
representation of propagators \cite{Schwinger}. We also apply double
Borel transformations over $p^2$ and $p^{\prime 2}$. The results after the Borel transformations are as follows:

 \bea
\label{e7326} \hat{I_0}(a,b,c)& =&i\frac{(-1)^{a+b+c}}{16
\pi^2\,\Gamma(a) \Gamma(b) \Gamma(c)}
(M_1^2)^{2-a-b} (M_2^2)^{2-a-c} \, {\cal U}_0(a+b+c-4,1-c-b)~, \nnb \\ \nnb \\
\hat{I_0}_\mu(a,b,c) &=&\hat{I_1}(a,b,c) p_\mu + \hat{I_2}(a,b,c) p'_\mu~,  \eea where
\begin{eqnarray}
 \hat{I_1}(a,b,c) &=& i \frac{(-1)^{a+b+c+1}}{16
\pi^2\,\Gamma(a) \Gamma(b) \Gamma(c)}
(M_1^2)^{2-a-b} (M_2^2)^{3-a-c} \, {\cal U}_0(a+b+c-5,1-c-b)~, \nonumber \\ \nonumber \\
\hat{I_2}(a,b,c) &=& i \frac{(-1)^{a+b+c+1}}{16 \pi^2\,\Gamma(a)
\Gamma(b) \Gamma(c)}
(M_1^2)^{3-a-b} (M_2^2)^{2-a-c} \, {\cal U}_0(a+b+c-5,1-c-b)~,
\end{eqnarray}
  and $M_1^2$ and $M_2^2$ are the Borel
parameters. The function ${\cal U}_0(\alpha,\beta)$ is  as follows:
\bea {\cal U}_0(a,b) = \int_0^\infty dy (y+M_1^2+M_2^2)^a y^b
\,exp\left[ -\frac{B_{-1}}{y} - B_0 - B_1 y \right]~, \nnb \eea
where \bea \label{e7328} B_{-1}& =& \frac{1}{M_1^2M_2^2}
\left[m_c^2M_1^4+m_b^2 M_2^4 + M_2^2M_1^2 (m_b^2+m_c^2
-q^2) \right] ~, \nnb \\
B_0 &=& \frac{1}{M_1^2 M_2^2} \left[ (m_b^2+m_c^2) M_1^2 + 2M_2^2 m_b^2
\right] ~, \nnb \\
B_{1} &=& \frac{m_b^2}{M_1^2 M_2^2}~. \eea
The circumflex of $\hat{I}$ in the equations is used for the result of integrals  after the  double Borel transformation.
After  lengthy calculations,  the following expressions for the two-gluon condensate contributions are obtained:

\begin{eqnarray}
\Gamma_{nonper}^{B_c^{PS}}&=&16 (6 m_b^3 (2 I_ 1 (1, 4, 1) + 3 I_ 1 (4, 1, 1) + I_ 2 (1, 4, 1)) -  6 m_b^2 m_c
       (I_ 1 (4, 1, 1) + I_ 2 (1, 4, 1))\nonumber\\& +&
   m_b (6 m_c^2 (2 I_ 1 (1, 1, 4) + I_ 2 (1, 1, 4)) -
      2 I_ 1 (1, 2, 2) + 6
          I_ 1 (1, 3, 1) + 6 I_ 1 (2, 1, 2) \nonumber\\& -& 2 I_ 1 (2, 2, 1) +
      6 I_ 1 (3, 1, 1) - I_ 2 (1, 2, 2) + 6 I_ 2 (1, 3, 1) + 3
          I_ 2 (2, 1, 2) - I_ 2 (2, 2, 1)) \nonumber\\& +&
   m_c (-6 m_c^2 I_ 2 (1, 1, 4) - 6 I_ 2 (1, 1, 3) + I_ 2 (1, 2, 2) - 3
          I_ 2 (2, 1, 2) + I_ 2 (2, 2, 1)))
   \end{eqnarray}

\begin{eqnarray}
\Gamma_{nonper}^{B_c^{V}}&=&-16 (-m_b (I (2, 1, 2) + I (2, 2, 1) - 12 m_c^2 I_ 1 (1, 1, 4) -
      6 I_ 1 (1, 2, 2) - 6 I_ 1 (1, 3, 1)\nonumber\\& +& 2
          I_ 1 (2, 1, 2) + 2 I_ 1 (2, 2, 1) - 6 I_ 1 (3, 1, 1) -
      6 m_c^2 I_ 2 (1, 1, 4) - 3 I_ 2 (1, 2, 2) - 6
          I_ 2 (1, 3, 1) + I_ 2 (2, 1, 2)\nonumber\\& +& I_ 2 (2, 2, 1)) +
   6 m_b^3 (2 I_ 1 (1, 4, 1) + 3
          I_ 1 (4, 1, 1) + I_ 2 (1, 4, 1)) -
   6 m_b^2 m_c (I_ 1 (4, 1, 1) + I_ 2 (1, 4, 1)) \nonumber\\& +& m_c (-6 m_c^2
          I_ 2 (1, 1, 4) - 6 I_ 2 (1, 1, 3) - 3 I_ 2 (1, 2, 2) +
      I_ 2 (2, 1, 2) + I_ 2 (2, 2, 1)))
   \end{eqnarray}

After applying the Borel transformation to  both physical and theoretical sides, we equate   the coefficients of the $(p.p^\prime)g_{\mu\nu}$ structure from both sides(physical  and QCD sides). The results related to the  sum rules for the corresponding form factors are found to  be:

\begin{eqnarray}\label{CoupCons-YbBB-Bpsoffshel}
g^{i}_{h_b B_c^{PS}B_c{V}}(q^2)&=&\frac{2(q^2-m_{i}^2)(m_b+m_c)}{f_{h_b}
f_{B_C^{PS}} f_{B_c^{V}}m^2_{B_C^{PS}}m_{h_b}m_{B_c^{V}}}
e^{\frac{m_{h_b}^2}{M^2}}e^{\frac{m_{(j)}^2}{{M^{\prime}}^2}}
\left[\frac{1}{4~\pi^2}\int^{s_0}_{4m_b^2}
ds\int^{s^{\prime}_0}_{(m_b+m_c)^2} ds^{\prime}
\rho^{i}(s,s^{\prime},q^2)\right.
\nonumber \\
&& \left. \theta
[1-{(f^{i}(s,s^{\prime}))}^2]e^{\frac{-s}{M^2}}e^{\frac{-s^{\prime}}
{{M^{\prime}}^2}}+\Pi^{i}_{nonper}\right],
\end{eqnarray}
 where; $i$ and $j$ are either $B_c^{PS}$ or $B_c^{V}$,  where ($i\neq j $).
\section{Numerical analysis}
In this study, we calculate  the form factor with both the $\bar{MS}$ and pole masses. The values  given in the Review of Particle Physics are
$\bar{m}_c(\bar{m}_c^2)=1.275\pm 0.025 $GeV and   $\bar{m}_b(\bar{m}_b^2)=4.18\pm 0.03 $GeV\cite{pdg}, which correspond to the pole
 masses $m_c=1.65\pm0.07$GeV and $m_b=4.78\pm0.06$GeV\cite{Wang, Ioffe}. A summary of the other input parameters are given in Table I.
\begin{center}
{\small TABLE
I: Values of pole masses of quarks  and decay constants  used in the calculation.}

{\small \bigskip}%
\begin{tabular}
[c]{ccccccc}\hline
$m_{B_{c}^{PS}}${\small \ \cite{pdg}} &$m_{B_{c}^{V}}${\small \ \cite{Wang}} & $m_{h_{b}}${\small \cite{pdg}}
&$f_{B_{C}^{V}}${\small \cite{Wang}} & $f_{B_{C}^{PS}}${\small \ \cite{Veliev}} & $f_{h_{b}}${\small \ \cite{Bashiry}}\\\hline
${\small 6.2745\pm0.0018}$ & ${\small 6.331\pm0.017}$ & ${\small 9.899.3\pm0.001}$ & ${\small 0.415\pm0.031}$ & ${\small 0.40\pm0.025}$ & ${\small 0.094\pm0.01}$\\\hline
\end{tabular}
\end{center}
 The sum rules contain  auxiliary parameters, namely Borel
mass parameters $M^{2},M^{\prime2}$ and continuum threshold ($s_{0}$ and
$s_{0}^{\prime}$). The standard criterion in QCD sum rules is that the physical
quantities are independent of the auxiliary parameters. Therefore, we search for
the intervals of these parameters so that our results are almost
insensitive to their variations. One more condition for the intervals of
the Borel mass parameters is the fact that the aforementioned intervals must
suppress the higher states, continuum and contributions of the highest-order operators.
 In other words, the sum rules for the form factors must
converge. As a result, we get $25GeV^{2}\leq M^{2}\leq30GeV^{2}$and
$20GeV^{2}\leq M^{\prime2}\leq25GeV^{2}$ for both $B_{c}^{PS} $ and $B_{c}%
^{V}$off-shell associated with the $h_b{B_{c}^{PS}B_{c}^{V}}$ vertex.

We depict the dependence of strong coupling constants on Borel parameters for
$B_{c}^{V}$ off-shell in Figs. 3 and 4) These figures indicate the weak
dependence of form factor of  $B_{c}^{V}$ off-shell in terms of the
Borel mass parameters in the chosen intervals. We find  stable
behavior of coupling constant in terms of the Borel mass parameters for the
$B_{c}^{PS}$ off-shell case and we find it unnecessary to show the other figures.

The continuum thresholds $s_{0}$ and $s_{0}^{\prime}$ are not arbitrary, but
correlated to the energy of the first excited state with the same quantum
number as the interpolating current. Thus, we choose the following regions for
the continuum thresholds in $s_{0}$ and $s_{0}^{\prime}$ channels:

\begin{equation}\label{s}
(m_{h_{b}%
}+0.4)^{2}\leq s_{0}\leq(m+0.6)^{2}
\end{equation}
 in $s$ channel for both off-shell cases,

\begin{eqnarray}
 &(m_{B_{C}^{PS}}+0.4)^{2}\leq s_{0}^{\prime}\leq(m_{B_{C}^{PS}}+0.6)^{2}& \label{sp1} \\
&(m_{B_{c}^{V}}+0.4)^{2}\leq s_{0}^{\prime}\leq(m_{B_{c}^{V}}+0.6)^{2}&\label{sp2}
\end{eqnarray}
for $B_{c}^{PS}$ and $B_{c}^{V}$ off-shell cases, respectively in
$s_{0}^{\prime}$ channel.

As a final remark, we should say that we follow the
standard procedure in the QCD sum rules where the continuum thresholds are
supposed to be independent of the Borel mass parameters and of $q^{2}$.
However, this standard assumption seems not to be  accurate, as mentioned in Ref.\cite{Lucha}.
\begin{figure}[h!]
\begin{center}
\includegraphics[width=12cm]{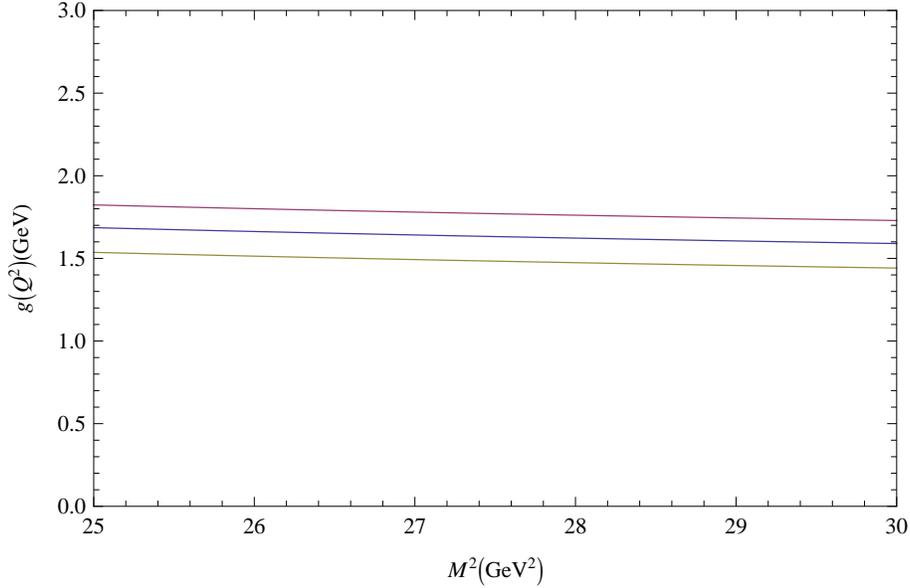}
\end{center}
\caption{$g^{B_c^{PS}}_{h_b B_c^V B_c^{PS}}(Q^2=5~GeV^2)$ as a function of
the Borel mass ${M}^2$. The continuum thresholds,
$s_0=(106.08,108.16,110.25)~GeV^2$, $s_0^{\prime}=(45.3, 46.66,48.3)~GeV^2$ and ${M^{\prime}}^2=20~GeV^2$
are used. The green, blue and purple lines are for minimum central and maximum values of $s_0$ and $s_0^\prime$ } \label{BMS}
\end{figure}

\begin{figure}[h!]
\begin{center}
\includegraphics[width=12cm]{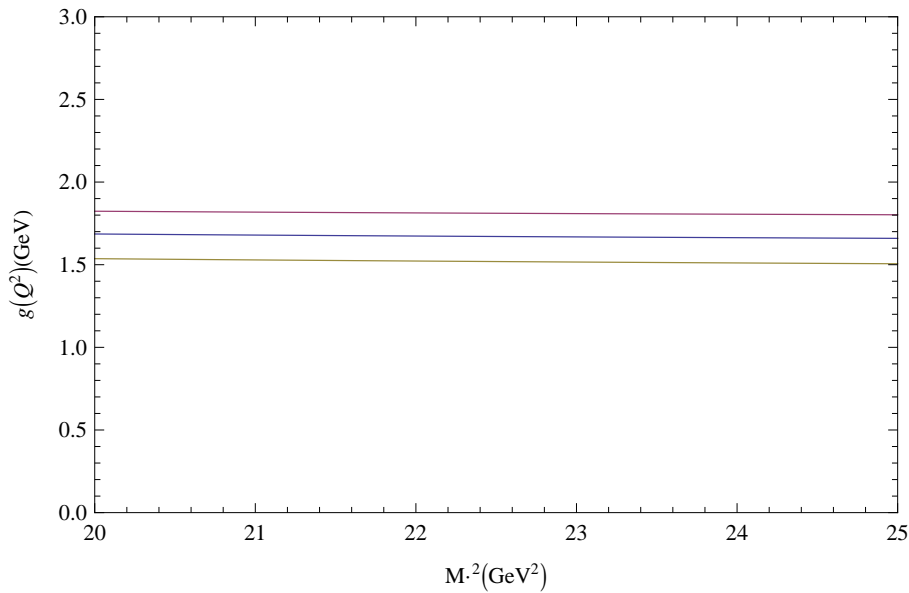}
\end{center}
\caption{$g^{B_c^{PS}}_{h_b B_c^V B_c^{PS}}(Q^2=5~GeV^2)$ as a function of
the Borel mass ${M}^2$. The continuum thresholds,
$s_0=(106.08,108.16, 110.25)~GeV^2$, $s_0^{\prime}=(45.3, 46.66,48.3)~GeV^2$ and ${M}^2=25~GeV^2$
are used. The green, blue and purple lines are for minimum central and maximum values of $s_0$ and $s_0^\prime$  } \label{BMSp}
\end{figure}

Our further numerical analysis shows that the dependence of the form factors on $q^{2}$ with the definite values of auxiliary parameters fits with the  following function:

\begin{equation}\label{fit}
g_{h_{b}B_{c}^{V}B_{c}^{PS}}^{i}(Q^{2})=A e^{ B Q^2}+C%
\end{equation}
where $Q^{2}=-q^{2}$, $i$ stands for $B_{c}^{PS}$ and $B_{c}^{V}$ off-shell
cases, and the value of $A, ~B$ and $C$ are shown in Table II.

By definition, the coupling constant is  the value of
$g_{h_{b}B_{C}^{V}B_{C}^{PS}}^{i}(Q^{2})$ at $Q^{2}=-m_{meson}^{2}$   \cite{Rodrigues}, where $m_{meson}$ is the mass of the on-shell mesons.

{\small TABLE II. Value of A, B and C for fit function for }$B_{c}^{PS}%
${\small \ and }$B_{c}^{V}${\small \ off-shell cases:}%
\[%
\begin{tabular}
[b]{|c|c|c|}\hline
& $B_{c}^{V}$ off-shell & $B_{c}^{PS}$ off-shell\\\hline
$A$ & $2.30 \pm0.50$& $2.43\pm 0.51$\\\hline
$B $& $0.035\pm 0.008$ & $0.033\pm 0.008$\\\hline
$C$ & $-0.31\pm 0.01$ & $-0.36\pm 0.11$\\\hline
\end{tabular}
\]
Substituting $Q^{2}=-m_{B_{c}^{PS}}^{2}$ and $Q^{2}=-m_{B_{c}^{V}}^{2}$ in Eq. (\ref{fit}), the $g_{h_{b}B_{C}^{V}B_{C}^{PS}}^{B_{C}^{PS}}=8.80\pm 2.84 GeV^{-1}$
and $g_{h_{b}B_{C}^{V}B_{C}^{PS}}^{B_{C}^{V}}=9.34\pm 3.12GeV^{-1}$ are
obtained for $B_{c}^{PS}$ and $B_{c}^{V}$ off-shell cases, respectively. The average value of the $g_{h_{b}B_{C}^{V}B_{C}^{PS}}$ strong coupling constant
is found to be
\begin{equation}
g_{h_{b}B_{C}^{V}B_{C}^{PS}}=(9.07\pm 2.93)GeV^{-1}%
\end{equation}

Note that  roughly $80\%$ of the  errors in our numerical calculation arise from the variation continuum  thresholds in intervals shown in Eqs. (\ref{s},\ref{sp1}) and \ref{sp2},   and remaining $20\%$ occure as a result of the   quark masses
 when one proceeds from the $\bar{MS}$ to the pole-scheme mass parameters,  the input parameters.

In conclusion, we calculate the strong coupling constant $g_{h_{b}B_{C}^{V}B_{C}^{PS}}$ using the three-point QCD sum rules. Our results show that the average value of the  strong coupling constant is $g_{h_{b}B_{C}^{V}B_{C}^{PS}}=(9.93\pm 2.7)GeV^{-1}$. Furthermore, the errors in our numerical calculations  depend on continuum threshold and  variation of the quark masses in different mass schemes.

\end{document}